\documentclass[11pt,a4paper,reqno]{amsart} 
\usepackage{amsmath, amsthm, amssymb,  mathrsfs}
\usepackage{graphicx}
\usepackage{float}
\newcounter{mnotecount}[section]

\renewcommand{\themnotecount}{\thesection.\arabic{mnotecount}}

\newcommand{\mnote}[1]
{\protect{\stepcounter{mnotecount}}$^{\mbox{\footnotesize
$
\bullet$\themnotecount}}$ \marginpar{
\raggedright\tiny\em
$\!\!\!\!\!\!\,\bullet$\themnotecount: #1} }

\def\be{\begin{equation}}
\def\ee{\end{equation}}
\def\bea{\begin{eqnarray}}
\def\eea{\end{eqnarray}}
\newcommand{\R}{\mathbb{R}}
\newcommand{\MM}{{\mathcal{M}}}
\begin{document}\date{2 September 2016}

\title{Non-diagonal four-dimensional cohomogeneity-one Einstein metrics in various signatures}

\author{Maciej Dunajski}
\address{Department of Applied Mathematics and Theoretical Physics\\
University of Cambridge\\ Wilberforce Road, Cambridge CB3 0WA, UK.
}
\email{m.dunajski@damtp.cam.ac.uk}

\author{Paul Tod}
\address{The Mathematical Institute\\
Oxford University\\
Woodstock Road, Oxford OX2 6GG\\ UK.
}
\email{tod@maths.ox.ac.uk}
\maketitle
\begin{abstract}
Most known four-dimensional cohomogeneity-one Einstein metrics are diagonal in the basis defined by the left-invariant one-forms, though some essentially non-diagonal ones are known. We consider the problem of
explicitly seeking non-diagonal Einstein metrics, and we find solutions which in some cases exhaust the possibilities. In particular we construct new examples of neutral signature
non--diagonal Bianchi type VIII Einstein metrics with self--dual Weyl tensor.
\end{abstract}

\section{Introduction}

A (pseudo) Riemannian four--manifold $(\MM, g)$
is said to have cohomogeneity--one
if it admits an isometry group $G$ acting transitively on
codimension-one surfaces in $\MM$.
Cohomogeneity-one Einstein metrics have been much studied both in the literature of general relativity, where they provide a simple generalisation of
the Friedman-Roberson-Walker cosmological models (see e.g. \cite{EM}), and in Riemannian geometry where they provide large classes of explicit solutions (see e.g. \cite{T1}). In the Riemannian case, the metric on the surfaces of homogeneity is inevitably positive-definite,
while in general relativity the surfaces of homogeneity are generally taken to be space-like when the metric is again definite (positive or negative according to convention).

When the surfaces of homogeneity are 3-dimensional, which is the case most-studied in relativity, the most general case has a 3-dimensional transitive group of isometries and these are classified by
the Bianchi classification of 3-dimensional Lie algebras, which has been an important part of mathematical cosmology for many years(see the historical account in \cite{Kplus}). We won't review the classification here,
for that see e.g. \cite{EM} or \cite{T1}, but we will use the language of the classification.
In \cite{DM} the authors exhibited an explicit Bianchi type VIII Einstein metric with self-dual Weyl tensor (a so-called `SD' solution) which is \emph{nondiagonal} in the sense that
the metric on the surfaces of homogeneity (which for brevity we shall call the \emph{spatial metric}) is not diagonal in the basis of left-invariant one-forms associated
with the symmetry, and cannot be diagonalised in this basis.
The metric has signature $(2,2)$ (or \emph{neutral signature}) so that the spatial metric is indefinite. It is known that an Einstein type VIII or IX metric with definite spatial metric may always be assumed to be diagonal
in the sense used here:
Locally ${\mathcal{M}}=\R\times G$, and
the group coordinates do not appear in the Einstein tensor.
The problem of finding  cohomogeneity--one Einstein metrics
therefore  reduces to solving a system of second-order ODEs on the spatial metric with independent variable that can conveniently be called \emph{time} regardless of the signature;
if the spatial metric is diagonalised at one time, then one of the Einstein constraints forces it to stay diagonal at all
times. This depends crucially on the field equations and does not necessarily hold
for Einstein equations with matter sources. This is made explicit in \cite{T1} for Riemannian signature and in \cite{EM} for many cases in relativistic cosmology. By an examination of
the proof one can see that real type IX metrics of any signature can be assumed to be diagonal without loss of generality, but if one wants nondiagonal type VIII metrics, one may find them in neutral signature. The interplay between diagonalisability and various signatures of $g$ is related to
the applicability of Sylvester's Law of Inertia, and can be seen as follows:
A general cohomogeneity--one metric takes the form
\[
g=dt^2+h_{ij}(t)\sigma^i\sigma^j,
\]
where the $\sigma^i$ are the left--invariant one forms on $G$ such that
\[
d\sigma^i+\frac{1}{2}C_{jk}^i\sigma^j\wedge\sigma^k=0.
\]
The matrix $n^{ij}=C_{km}^i\epsilon^{kmj}$ is symmetric for type VIII. If $h_{ij}$ is positive (or negative) definite, then one can diagonalise $h$ and $n$  simultaneously by changing the basis
$\sigma^i$. This is not always true if $g$ is neutral, so that $h$ has indefinite signature.

 Nondiagonal Einstein metrics of other Bianchi types are known (see e.g. \cite{luk} for a Lorentzian vacuum solution and \cite{CT1} and \cite{CT2} for a variety of solutions). It is worth noticing that all
types apart from VIII and IX have an isometry group with a 2-dimensional Abelian subgroup, and when this group acts orthogonally-transitively the Einstein vacuum equations in any signature are
integrable and can be solved by the twistor methods of Mason and Woodhouse and collaborators, \cite{MW}, following \cite{wit} and \cite{W}\footnote{In particular all the examples in \cite{CT1} and \cite{CT2} do have orthogonally transitive $G_2$ actions, and this
is in a sense the reason why the integrations there lead to
Painlev\'e equations.}.
\vskip5pt
In this paper we set out to find real non-diagonal cohomogeneity-one Einstein metrics. We begin in Section 2 with the simplest case, type I, which has a three-dimensional Abelian isometry group. This case is fairly well-known
but the calculation highlights the interaction between the signature and the variety of canonical forms. Some new examples are given by formulae
(\ref{met2}) and (\ref{met5}).
In Section 3 we show how to obtain nondiagonal real type VIII from complex type IX, both diagonal and nondiagonal. 
In particular the class class of metrics (\ref{met11}) contains the 
self--dual Einstein example \cite{DM} with $\Lambda\neq 0$ as special case,
and the three--parameter family of Ricci--flat metrics (\ref{met44}) 
generalises the Eguchi--Hanson solution.
In Section 4 we briefly consider other types. Where these have orthogonally transitive subgroups of the isometry group they can be solved by twistor methods so we consider direct integration of some examples which do not have orthogonally transitive subgroups of the isometry group. An example of a non--diagonal, self-dual Ricci--flat Bianchi II metric is given by 
(\ref{met55}).

\section{Type I}
The integrations are particularly simple in this case, but this enables one to see how the different metrics arise from the classification of minimum polynomials of a relevant matrix, and introduces the methods needed later
for the more complicated types.
\subsection{The vacuum and Einstein equations}
For type I, the invariant one-forms are all exact so that one may choose three spatial coordinates $(x^1,x^2,x^3)$ and the metric can be taken to be
\be\label{four_metric}g=dt^2+h_{ij}(t)dx^idx^j.\ee
We may calculate the curvature in an unsophisticated manner. First the Christoffel symbols are given by
\[\Gamma^0_{00}=\Gamma^0_{0i}=\Gamma^i_{00}=\Gamma^i_{jk}=0\]
while
\begin{eqnarray*}
\Gamma^0_{ij}&=&-\frac12\dot{h}_{ij}\\
\Gamma^i_{0j}&=&\frac12 h^{ik}\dot{h}_{jk},\end{eqnarray*}
with overdot for $d/dt$. Then with the definition of the Riemann tensor as
\[R^{a}_{\;bcd}=-2\Gamma^a_{b[c,d]}-2\Gamma^e_{b[c}\Gamma^a_{d]e},\]
we find
\bea\label{r1}
R_{\;i0j}^{0}&=&-\frac12 \ddot{h}_{ij}+\frac14 h^{km}\dot{h}_{im}\dot{h}_{kj},\nonumber\\
R_{\;ijk}^{0}&=&0,\\
R_{\;jkl}^{i}&=&\frac14 h^{im}\dot{h}_{jk}\dot{h}_{ml}-\frac14 h^{im}\dot{h}_{jl}\dot{h}_{mk}.\nonumber
\eea
(It follows rapidly from here that self-duality or anti-self-duality of the Riemann tensor at once implies that the curvature entirely vanishes.)

\medskip

For the Ricci tensor we obtain
\bea\label{r2}
R_{00}&=&-\frac12 h^{ij}\ddot{h}_{ij}+\frac14 h^{km}\dot{h}_{kj}h^{ij}\dot{h}_{im},\nonumber\\
R_{0i}&=&0,\\
R_{ij}&=&-\frac12 \ddot{h}_{ij}+\frac12 h^{km}\dot{h}_{im}\dot{h}_{kj}-\frac14(h^{mn}\dot{h}_{mn})\dot{h}_{ij}.\nonumber
\eea
For the vacuum equations, double the last of (\ref{r2}), multiply by $-h^{ki}$ and adopt a matrix notation:
\be\label{mat5}
h^{-1}\ddot{h}-h^{-1}\dot{h}h^{-1}\dot{h}+\frac12 \mbox{tr}(h^{-1}\dot{h})h^{-1}\dot{h}=0.\ee
To simplify this introduce $V$ by
\[2\dot{V}/V=\mbox{tr}(h^{-1}\dot{h})=\frac{d}{dt}\log(\mbox{det}h),\]
so that $|\mbox{det}h|=V^2$, then (\ref{mat5}) becomes
\[
d(Vh^{-1}\dot{h})/dt=0\]
which integrates at once to
\be\label{v1}Vh^{-1}\dot{h}=M=\mbox{constant},\ee
or equivalently

\be\label{e1}
\dot{h}=V^{-1}hM.
\ee
Take the trace of (\ref{v1}) to find
\[\mbox{tr}(M)=2\dot{V}\]
so that $\ddot{V}=0$ and $V$ is linear in $t$.

Go back to the first of (\ref{r2}) to get another vacuum equation which becomes
\be\label{e2}\mbox{tr}(M^2)-(\mbox{tr}M)^2=0,\ee
and which constrains $M$. Since $h$ and $\dot{h}$ are both symmetric, $M$ is also constrained by (\ref{e1}). Clearly the 4-metric is flat if $M=0$.

At this point it is worth noting the Einstein equations with $\Lambda\neq 0$. Suppose these are
\[R_{ij}=\Lambda h_{ij},\;\;\;R_{00}=\Lambda\]
then
\be\label{e3}d(Vh^{-1}\dot{h})/dt=-2\Lambda V\mathbb{I}\ee
with $2\dot{V}/V=\mbox{tr}(h^{-1}\dot{h})$ as before. Now however, using the $(00)$ equation
\[\ddot{V}=-3\Lambda V.\]
Solve this with $V(0)=1,\dot{V}(0)=0$ to find
\[V=\cos(3Ht)\mbox{  or  }\cosh(3Ht)\]
when $H^2=\Lambda/3$ or $-\Lambda/3$ respectively. Integrate (\ref{e3}) once to find
\[Vh^{-1}\dot{h}=M+\frac23\dot{V}\mathbb{I} ,\]
with constant $M$. Introduce $k=V^{-2/3}h$ then $k$ satisfies
\[Vk^{-1}\dot{k}=M\]
which is (\ref{e1}) back again. Thus the solutions with $\Lambda\neq 0$ are readily obtained from the solutions with $\Lambda=0$.
\subsection{Classification of vacuum solutions}
The solutions depend on the canonical forms of $h$ and $M$.
\begin{enumerate}
 \item
 If $h$ is definite (positive or negative) then, at any fixed instant say $t_0$ of $t$, it can be diagonalised with an orthogonal matrix and reduced to the identity by a diagonal matrix. At this instant, (\ref{e1})
 forces $M$ to be symmetric so it too can be diagonalised by an orthogonal matrix preserving $h$. But $M$ is constant so now $h$ is diagonal for all time. These are the Kasner solutions.
 \item If $h$ is indefinite, it can still be reduced at any fixed instant to a diagonal matrix whose entries are $\pm 1$. For simplicity suppose this is
 \[h(t_0)=\eta:=\mbox{diag}(1,1,-1),\]
 since the other possibility not so far covered follows by $x^j\rightarrow ix^j$. We are at liberty to perform the transformation
 \[h\rightarrow\hat{h}=L^ThL,\;\;M\rightarrow\hat{M}=L^{-1}ML,\]
 where $L$ is a three-dimensional Lorentz matrix. By (\ref{e1}), at time $t_0$ the matrix $S:=\eta M$ is symmetric.

 To classify canonical forms of $M$ we consider the eigenvalue equation
 \[M_{\alpha}^{\;\;\beta}X^\alpha=\lambda X^\beta,\]
 where Greek indices are 3-dimensional Lorentz indices (and sometimes omitted) and $h=\eta_{\alpha\beta}$ is the Lorentz metric. There are three possibilities:

 \medskip

 \begin{enumerate}
 \item
 If there are three distinct real $\lambda$ then $M$ can be diagonalised and we obtain Kasner again.
 \item
 If there are repeated real roots we need to consider the minimum polynomial $m(x)$ of $M$. Taking account of (\ref{e2}) there are three nontrivial cases namely
 \[(x-\lambda)(x-\mu)^2,\;x^2,\;x^3\]
 (we don't need to consider $(x-\lambda)^k$
 for nonzero $\lambda$ as this would violate (\ref{e2}), and we don't need to consider $m(x)=x$ as this gives flat space).

 \medskip

 \begin{enumerate}
  \item $m(x)=(x-\lambda)(x-\mu)^2$ with $\lambda,\mu$ distinct. Necessarily $\lambda$ and $\mu$ are real and (\ref{e2}) requires
  \[\mu(\lambda+2\mu)=0,\]
  so in particular $\lambda=0$ implies $\mu=0$ and we have the next case so w.l.o.g. $\lambda\neq 0$; also $\lambda=\mu$ implies $\lambda=0$ and is therefore ruled out.

  There is a basis of vectors $X,Y,Z$ (omitting the indices) with
  \[MX=\lambda X,\;\;MY=\mu Y,\;\;MZ=\mu Z+Y.\] In terms of the matrix $S=S_{\alpha\beta}$ we deduce
  \[\lambda X_\alpha Y^\alpha=S_{\alpha\beta}X^\alpha Y^\beta=\mu X_\alpha Y^\alpha\]
  so $X_\alpha Y^\alpha=0$ and similarly $X_\alpha Z^\alpha=0$. Next
  \[\mu Y_\alpha Z^\alpha=S_{\alpha\beta}Y^\alpha Z^\beta=\mu Y_\alpha Z^\alpha+Y_\alpha Y^\alpha \]
  so that $Y$ is a null vector.

  Since we are free to add multiples of $Y$ to $Z$ and perform Lorentz transformations, we can assume
  \[X=(1,0,0)^T,\;\;Y=(0,1,1)^T,\;\;Z=(0,A,-A)^T,\]
  for some nonzero, real $A$ and we find

  \[
M=\left(\begin{array}{ccc}
           \lambda & 0 & 0\\
          0 & \mu+\nu & -\nu\\
           0 & \nu & \mu-\nu\\
\end{array}\right).
\]
  where $\nu=1/(2A)$, and still $\mu(\lambda+2\mu)=0$. To find the metrics, take

  \[h=\left(\begin{array}{ccc}
           \alpha & \beta & \gamma\\
          \beta & \delta & \epsilon\\
           \gamma & \epsilon & \zeta\\
\end{array}\right),
\]
with $h(0)=\eta$ and substitute into (\ref{e1}). Symmetry of $\dot{h}$ requires
  \[(\mu+\nu-\lambda)\beta+\gamma\nu=0=-\beta\nu+\gamma(\mu-\nu-\lambda),\;\;\nu(2\epsilon+\delta+\zeta)=0.\]
  Since $\lambda-\mu\neq 0$, the first pair imply $\beta=\gamma=0$ and the vacuum equations reduce to
  \bea\label{vac1}
\dot{\alpha}&=&\frac{1}{V}\lambda\alpha\nonumber\\
\dot{\delta}&=&\frac{1}{V}(\delta(\mu+\nu)+\epsilon\nu)\nonumber\\
 \dot{\epsilon}&=&\frac{1}{V}(-\delta\nu+\epsilon(\mu-\nu))\\
 \dot{\zeta}&=&\frac{1}{V}(-\epsilon\nu+\zeta(\mu-\nu))\nonumber\\
 \dot{V}&=&\frac12(\lambda+2\nu)\nonumber
 \eea
  There are two cases:
  \begin{enumerate}
  \item $\mu=0$ when
\be\label{met1}
h=(1+t/t_0)^2dx^2+dy^2-dz^2+\frac{2\nu}{\lambda}\log(1+t/t_0)(dy-dz)^2\ee
The resulting 4-metric (\ref{four_metric}) is a pp-wave (see e.g. \cite{exact_sol})).
 \item
 $\lambda+2\mu=0$ when
  \be\label{met2}
  h=e^{-2\mu t}dx^2+e^{\mu t}((dy^2-dz^2)+\nu t(dy-dz)^2).\ee

  \end{enumerate}

  \item
 $m(x)=x^2$: this follows from the previous case by setting $\lambda=\mu=0$ in $M$, and in fact in (\ref{met2}), but the metrics turn out to be flat.
%

 \item $m(x)=x^3$: there will be a basis $X,Y,Z$ with
 \[MX=0,\;\;MY=X,\;\;MZ=Y,\]
 and we are free to perform the changes
 \[Z\rightarrow Z+\alpha X+\beta Y,\;\;Y\rightarrow Y+\beta X.\]
 By similar considerations to those used above, we deduce
 \[X_\alpha X^\alpha=0=X_\alpha Y^\alpha=X_\alpha Z^\alpha -Y_\alpha Y^\alpha\]
 so, using Lorentz transformations and the available freedom, w.l.o.g.
 \[X=(0,1,1)^T,\;\;Y=(A,0,0)^T,\;\;Z=(C,D,-D)^T,\]
 with $AD\neq 0$. We can deduce $S$ and from it calculate $M$:

  \[
M=\left(\begin{array}{ccc}
           0 & \lambda & -\lambda\\
          \lambda & 0 & 0\\
           \lambda & 0 & 0\\
\end{array}\right),
\]
  with $\lambda\neq 0$. Note $\mbox{tr}M=0$ so w.l.o.g. $V=1$, and taking
  \[h=\left(\begin{array}{ccc}
           \alpha & \beta & \gamma\\
          \beta & \delta & \epsilon\\
           \gamma & \epsilon & \zeta\\
\end{array}\right),
\]
with $h(0)=\eta$, substitute into (\ref{e1}). Symmetry of $\dot{h}$ forces

\[\alpha=\delta+\epsilon=-\zeta-\epsilon,\;\;\beta+\gamma=0,\]
reducing the vacuum equations to the system
\bea\label{vac2}
\dot{\alpha}&=&0\nonumber\\
\dot{\beta}&=&\lambda\alpha\nonumber\\
 \dot{\delta}&=&\lambda\beta\\
 \dot{\epsilon}&=&-\lambda\beta\nonumber\\
 \dot{\zeta}&=&-\lambda\gamma\nonumber
 \eea
  which are solved by
  \[\alpha=1,\;\;\beta=-\gamma=\lambda t,\;\;\delta=1+\frac12\lambda^2t^2,\;\;\epsilon=-
  \frac12\lambda^2t^2,\;\;\zeta=-1+\frac12\lambda^2t^2.\]
  The metric is
  \be\label{met4}
  g=dt^2+dx^2+dy^2-dz^2+2\lambda tdx(dy-dz)+\frac12 \lambda^2t^2(dy-dz)^2.\ee
  For this metric, the components of the Riemann tensor as in (3)-(5) are constant (and not all zero) but the resulting space-time is not homogeneous as the Riemann tensor is not parallel\footnote{This is the metric found in \cite{CT2} equation (4.10)}.
 \end{enumerate}

 \item
 If there is one real root, say $\lambda$ and a complex conjugate pair, say $\mu\pm i\nu$ with real $\mu,\nu$ (and $\nu\neq 0$ or we are in an earlier case) then there are real Lorentz vectors $X,U,V$ with
 \[MX=\lambda X,\;\;M(U+iV)=(\mu+i\nu)(U+iV).\]
 With the aid of the symmetric matrix $S$ we can write these equations as
 \[S_{\alpha\beta}X^\beta=\lambda X_\alpha,\;\;S_{\alpha\beta}(U^\beta+iV^\beta)=(\mu+i\nu)(U_\alpha+iV_\alpha),\]
 and deduce
 \[X_\alpha U^\alpha=0=X_\alpha V^\alpha=U_\alpha U^\alpha+V_\alpha V^\alpha.\]
 By rotating $U$ and $V$ in the 2-plane they span:
 \[(U^\alpha+iV^\alpha)\rightarrow e^{i\phi}(U^\alpha+iV^\alpha)\]
 we can arrange that $U_\alpha V^\alpha=0$. Now they cannot both be null so one, say $U$ is space-like and then the other, $V$, is time-like. Since $X$ is orthogonal to both, it must be space-like and
 we have an orthogonal triad which we can suppose to be normalised. Choose
 \[X=(1,0,0)^T, \;\; U=(0,1,0)^T, \;\;V=(0,0,1)^T\]
 then
 \[
M=\left(\begin{array}{ccc}
           \lambda & 0 & 0\\
          0 & \mu & -\nu\\
           0 & \nu & \mu\\
\end{array}\right).
\]
 We need to impose (\ref{e2}), which implies
 \[2\lambda\mu+\mu^2+\nu^2=0.\]
 We know that $\nu\neq 0$ so this also forces $\mu\neq 0$ and we can solve for $\lambda$:
 \[\lambda=-\frac{(\mu^2+\nu^2)}{2\mu}.\]

 \medskip

 To find the metric explicitly, first parametrise $h$:
 \[h=\left(\begin{array}{ccc}
           \alpha & \beta & \gamma\\
          \beta & \delta & \epsilon\\
           \gamma & \epsilon & \zeta\\
\end{array}\right),
\]
with $h(0)=\eta$ and substitute into (\ref{e1}). Symmetry forces $\beta=\gamma=\zeta+\delta=0$ and leaves the system
\bea\label{vac4}
\dot{\alpha}&=&\frac{1}{V}\lambda\alpha\nonumber\\
\dot{\delta}&=&\frac{1}{V}(\mu\delta+\nu\epsilon)\\
 \dot{\epsilon}&=&\frac{1}{V}(-\nu\delta+\mu\epsilon)\nonumber\\
 \dot{V}&=&\frac{1}{2}(\lambda+2\mu)\nonumber
 \eea
 \begin{enumerate}
 \item If $\lambda+2\mu\neq 0$ then
 \[V=1+\frac{t}{t_0},\;\;\alpha=(1+\frac{t}{t_0})^{2p},\;\;\delta+i\epsilon=(1+\frac{t}{t_0})^{2(q+ir)}\]
 with
 \[p=\frac{\lambda}{\lambda+2\mu},\;\;q+ir=\frac{(\mu-i\nu)}{\lambda+2\mu},\;\;t_0=\frac{2}{\lambda+2\mu}.\]
 Note that
 \[p+(q+ir)+(q-ir)=1=p^2+(q+ir)^2+(q-ir)^2,\]
 so that this metric is essentially a real slice of a complex Kasner metric. This family of metrics appears in \cite{CT2}
as \emph{Harrison metrics}.

 \item If $\lambda+2\mu=0$ then also $\nu=\pm\mu\sqrt{3}$ and there is just one free parameter. We can assume $V=1$ without loss of generality and then
 \be\label{met5}\alpha=e^{-2\mu t},\;\;\delta=e^{\mu t}\cos(\nu t),\;\;\epsilon=-e^{\mu t}\sin(\nu t).\ee
   \end{enumerate}
  \end{enumerate}
\end{enumerate}
We have found all type I vacuum metrics of all signatures, of which (\ref{met2}) and (\ref{met5}) seem to be new. All type I Einstein metrics with nonzero $\Lambda$ can then be found by the method at the end of section 2.1.

\section{Types VIII and IX}
%
As noted in the Introduction, real type IX Einstein metrics of any signature can be assumed to be diagonal without loss of generality. Similarly type VIII Einstein with positive or negative definite spatial metric can be
  assumed diagonal without loss of generality. Thus for nondiagonal real Einstein metrics one should consider neutral signature type VIII.
\subsection{Nondiagonal type VIII from diagonal type IX}
 It is possible to obtain all real analytic nondiagonal definite or indefinite type VIII examples by taking real slices of complex type IX solutions. This can be seen as follows: suppose the left-invariant one-forms are
 $\Sigma_i$ for type IX and $\sigma_i$ for type VIII, so that
\[d\Sigma_1=\Sigma_2\wedge\Sigma_3, \quad
d\Sigma_2=\Sigma_3\wedge\Sigma_1,\quad
d\Sigma_3=\Sigma_1\wedge\Sigma_2\]
and
 \be\label{type_8}d\sigma_1=\sigma_2\wedge\sigma_3,\;\;d\sigma_2=-\sigma_3\wedge\sigma_1,\;\;d\sigma_3=\sigma_1\wedge\sigma_2,\ee
 then given a real analytic
 type VIII metric $g$ set
 \be\label{s1}\sigma_1=i\Sigma_1,\;\;\sigma_2=\Sigma_2,\;\;\sigma_3=-i\Sigma_3\ee
 which, following \cite{kam}, we shall call `Kamada's choice', to obtain a complex type IX metric; either this can be diagonalised at any choice of time and will then remain diagonal, so that
 the original type VIII metric is defined on a real slice of
 this complex (and diagonal) type IX metric, or it cannot be so diagonalised in which case there are different solutions (that we find below).

 However an explicit general solution of the vacuum equations for
 type IX metrics is not known (and, being chaotic, is never likely to be known) so to find explicit solutions we impose an extra condition, namely self-dual
 or anti-self-dual (SD or ASD) Weyl tensor. Now
 we may follow the method of \cite{T1}, at least for diagonal type IX.

 With the conventions of \cite{T1},
 write the type IX metric as
 \be\label{m9}g=w_1w_2w_3dT^2+\frac{w_2w_3}{w_1}\Sigma_1^2+\frac{w_3w_1}{w_2}\Sigma_2^2+\frac{w_1w_2}{w_3}\Sigma_3^2.\ee
 In \cite{T1} the SD Einstein equations are solved for this metric. In brief, the Levi-Civita connection is coded into three connection variables $(A_1,A_2,A_3)$ obtained as first-derviatives of the $w_i$, and the first-order
 system for the $A_i$ is reduced to a Painlev\'e VI equation for the variable $x=\frac{A_1-A_2}{A_3-A_2}$.

 We shall find nondiagonal real type VIII solutions by imposing a particular set of reality conditions, different from Kamada`s choice, and we shall show in an Appendix that these two choices are the only two choices, up to
 an appropriate equivalence.

 We set
 \be\label{ch1}\Sigma_1=-\overline{\Sigma}_1=-i\sigma_1,\;\;\Sigma_2=\overline{\Sigma}_3=\frac{1}{\sqrt{2}}(\sigma_2+i\sigma_3)\ee
 for real $\sigma_i$, then relations (\ref{type_8}) hold, as required.

 To obtain a real metric, we have
 \be\label{m10}h=A\Sigma_1^2+B\Sigma_2^2+C\Sigma_3^2=-A\sigma_1^2+\frac12(B+C)(\sigma_2^2-\sigma_3^2)+i(B-C)\sigma_2\sigma_3,\ee
 which will be real if $A$ is real and $B=\overline{C}$. Returning to (\ref{m9}) we see that these conditions require $w_1$ real
 and $w_2=\pm\overline{w}_3$ (w.l.o.g. we take $w_2=\overline{w}_3$ as the other choice simply changes the overall sign of the metric).

 Following the conventions of \cite{T1}, we find that the functions $A_i$ that encode the connection coefficients must satisfy
$A_1\in\mathbb{R},\;\;A_2=\overline{A}_3.$ Introduce
\[
z=\frac{x}{x-1}=\frac{A_1-A_3}{A_1-A_2},
\]
in terms of $x(T)$ (with $T$ related to $t$ in (\ref{four_metric}) by $dt^2=w_1w_2w_3dT^2$) then the reality conditions force $z\overline{z}=1$ or equivalently $x+\overline{x}=1$. Substituting
$x=\frac12+iy$ into (4.8) of \cite{T1} we obtain an equation for $y$:
\[
\frac{d}{dT}\left((y')^{-3/2}y''\right)=\frac12 (y')^{3/2}
\frac{\frac34-y^2}{(\frac14+y^2)^2}, \quad\mbox{where}\quad '=d/dT.
\]
This will have real solutions for $y$, which will in turn give appropriate $z$.
To find the metric, continue as in \cite{T1}, and there will be new SD vacuum metrics in this class.

\subsection{Nondiagonal type VIII from non-diagonalisable type IX}
While every real symmetric $3\times 3$ matrix can be diagonalised by conjugation with a real orthogonal matrix, this is not true for complex symmetric $3\times 3$ matrix.
Consequently a real analytic type VIII metric may complexify to a nondiagonalisable type IX metric and then it will not lie in the previous class.

There are two relevant canonical forms of nondiagonalisable complex symmetric $3\times 3$ matrix, which are distinguished by the minimum polynomial. If the minimum poynomial has distinct roots
then the matrix is diagonalisable so for nondiagonalisability there must be repeated roots and the cases are:

\begin{enumerate}
 \item Minimum polynomial $(x-\lambda)(x-\mu)^2$ with $\lambda\neq\mu$, which leads to a spatial metric of the form

 \[\left(\begin{array}{ccc}
 \lambda&0&0\\
            0 & \mu+\nu&i\nu\\
           0& i\nu&\mu-\nu\\
 \end{array}\right)\]

 \item  Minimum polynomial $(x-\lambda)^3$ which leads to spatial metric

 \[\left(\begin{array}{ccc}
 \lambda-i\mu& \mu &\nu\\
            \mu & \lambda+i\mu&i\nu\\
           \nu& i\nu&\lambda\\
 \end{array}\right)\]

 \end{enumerate}
From these by the Kamada choice we obtain two type VIII metrics. Choose the invariant one-forms to satisfy (\ref{type_8}) and parametrise the metrics as
\begin{enumerate}
\item
\be
\label{first11}
g=dt^2-A^2\sigma_1^2+(\sigma_2-\sigma_3)((B+C)\sigma_2+(B-C)\sigma_3),
\ee
with $AB\neq 0$.
\item
\be\label{second22}
g=dt^2-(\sigma_1+\sigma_2)((A^2+B)\sigma_1-(A^2-B)\sigma_2+2C\sigma_3)-A^2\sigma_3^2,
\ee
with $A\neq 0$.
\end{enumerate}
These are evidently real for $A,B,C$ real, and both have neutral signature. We will solve the Einstein equations in the two cases, sometimes completely and sometimes reducing to a second-order linear ODE which can be
regarded as integrable.
\subsubsection{Solving the Einstein equations for the metric
(\ref{first11})}
The Einstein equations $R_{ab}=\Lambda g_{ab}$
for (\ref{first11}) are
\begin{eqnarray}\label{ee1}
-\frac{\ddot{A}}{A}-\frac{\ddot{B}}{B}+\frac{\dot{B}^2}{2B}&=&\Lambda\nonumber\\
\frac{\ddot{A}}{A}+\frac{\dot{A}\dot{B}}{AB}-\frac{A^2}{2B^2}&=&-\Lambda\\
\dot{P}+\frac{\dot{A}}{A}P-C\left(\frac{2}{A^2}-\frac{1}{B}\right)&=&0\nonumber\\
-\frac{\ddot{B}}{2B}-\frac{\dot{A}\dot{B}}{2AB}-\frac{A^2}{2B^2}+\frac{1}{B}&=&\Lambda\nonumber
\end{eqnarray}
where
\[P=\frac{1}{2B}(B\dot{C}-C\dot{B}).\] Only the third of (\ref{ee1}) contains $C$ so we leave it until last. From the others, by elimination of second derivatives, we obtain
\be\label{ham}
H:=\frac{\dot{B}^2}{2B^2}+2\frac{\dot{A}\dot{B}}{AB}+\frac{A^2}{2B^2}-\frac{2}{B}+2\Lambda=0,
\ee
which is the Hamiltonian constraint (and is conserved by virtue of the others).

Define
\[q:=\frac{\dot{B}^2}{A^2B}-\frac{1}{B},\]
then $\dot{q}=0$ by virtue of (\ref{ee1}), so $q=c_1$. Evidently the system (\ref{ee1}) is now equivalent to this and (\ref{ham}), together with the equation for $P$, and
we deduce
\be\label{a}
A^2=\frac{\dot{B}^2}{1+c_1B}
\ee
for constant $c_1$. Use this to eliminate $A$ from (\ref{ham}) to find
\[\ddot{B}+\frac{(2-c_1B)}{4B(1+c_1B)}\dot{B}^2=1-\Lambda B.\]
This integrates to give, if $c_1\neq 0$
\[\dot{B}^2=F(B)\] with
\be\label{b}
F(B):=\frac{2}{c_1^3B}\left(-\frac{2\Lambda}{3}(1+c_1B)^3+2(c_1+2\Lambda)(1+c_1B)^2+2(c_1+\Lambda)(1+c_1B)+c_2(1+c_1B)^{3/2}\right),
\ee
with a second constant of integration $c_2$, while if $c_1=0$
\be\label{b2}
\dot{B}^2=F(B):=\frac{c_2}{B}+B-\frac{2\Lambda}{3}B^2.\ee
Now go back to (\ref{ee1}) to find $C$. First note
\[AP=\frac{\dot{B}^2}{2(1+c_1B)^{1/2}}\left(\frac{dC}{dB}-\frac{C}{B}\right)=\frac{BF(B)}{2(1+c_1B)^{1/2}}\frac{d}{dB}\left(\frac{C}{B}\right),\]
so that the third of (\ref{ee1}) is
\be\label{c}
\frac{d}{dB}\left(\frac{BF}{2(1+c_1B)^{1/2}}\frac{d}{dB}\left(\frac{C}{B}\right)\right)=\frac{C}{B}\frac{1}{(1+c_1B)^{1/2}}\left(\frac{2B(1+c_1B)}{F}-1\right).
\ee
This is a second-order linear ODE for $C$ which we can suppose has been
solved, the solution incorporating two more constants $c_3,c_4$. The metric is
\[g=dt^2-A^2\sigma_1^2+(\sigma_2-\sigma_3)((B+C)\sigma_2+(B-C)\sigma_3),\]
so use $B$ as time-coordinate to obtain a solution depending on $c_1,...,c_4$ and $\Lambda$:
\be\label{met11}
g=\frac{dB^2}{F(B)}-\frac{F(B)}{1+c_1B}\sigma_1^2+B(\sigma_2^2-\sigma_3^2)+C(B)(\sigma_2-\sigma_3)^2,
\ee
where we are assuming $F(B)$ and $C(B)$ are known.

\medskip

In the special case $c_1=c_2=0$ we have $F=B-\frac{2\Lambda}{3}B^2$ and (\ref{c}) has the general solution
\[C=c_3\frac{B^2(1-\Lambda B/6)}{(1-2\Lambda B/3)^2}+c_4\frac{8\Lambda^2B^2-8\Lambda B+3}{(1-2\Lambda B/3)^2B},\]
which, with $c_4=0$, is recognisable as the solution in equation (7.5) of \cite{DM}. Thus that particular nondiagonalisable type VIII Einstein metric complexifies to a nondiagonalisable type IX metric.
\medskip

There is another simple special case: $c_1=0=\Lambda$. Then
$F(B)= B+c_2B^{-1}$,
and (\ref{c}) has the general solution
\[C=\frac{c_3B+3c_4c_2B^2+c_4B^4}{c_2+B^2},\]
with constants of integration $c_3,c_4$.
 With $B=r^2/4$ the metric can be written
\be\label{met44}
g=\frac{dr^2}{1+c_5/r^4}-\frac{r^2}{4}(1+c_5/r^4)\sigma_1^2+\frac{r^2}{4}(\sigma_2^2-\sigma_3^2)+\left(\frac{4c_3}{r^2}+3c_4c_5+c_4r^4\right)(1+c_5/r^4)^{-1}(\sigma_2-\sigma_3)^2.\ee
The Riemann tensor for this metric is SD iff $c_4=0$.
This metric
bears some resemblance to the Eguchi-Hansen metric but it seems to be new.

\subsubsection{Solving the Einstein equations for the
(\ref{second22})} The metric is
\[g=dt^2-(\sigma_1+\sigma_2)((A^2+B)\sigma_1-(A^2-B)\sigma_2+2C\sigma_3)-A^2\sigma_3^2,\]
with $A\neq 0$.
Choose the basis of forms to be
\[\theta^0=dt,\;\;\theta^1=A\sigma_3,\;\;\theta^2=\sigma_1+\sigma_2,\;\;\theta^3=\frac12(A^2+B)\sigma_1-\frac12(A^2-B)\sigma_2+C\sigma_3,\]
so that
\[g_{00}=-g_{11}=-g_{23}=1.\]
Now calculate the Einstein equations as
\bea
3\frac{\ddot{A}}{A}&=&-\Lambda\\
\frac{\ddot{A}}{A}+2\frac{\dot{A}^2}{A^2}-\frac{1}{2A^2}&=&-\Lambda\\
\ddot{C}-\frac{\dot{A}\dot{C}}{A}-2C\left(\frac{\ddot{A}}{A}+\frac{1}{A^2}\right)&=&0\label{cc}\\
\ddot{B}-\frac{\dot{A}\dot{B}}{A}-2B\left(\frac{\ddot{A}}{A}+\frac{1}{A^2}\right)&=&6\frac{C^2}{A^4}+\left(\frac{\dot{C}}{A}-2\frac{\dot{A}C}{A^2}\right)^2\label{bb}
\eea
Eliminate $\ddot{A}$ from the first pair to obtain
\[\frac{\dot{A}^2}{A^2}-\frac{1}{4A^2}=-\Lambda/3,\]
which is the Hamiltonian constraint. Solutions are
\be\label{aa}
\mbox{For  }\Lambda=-3H^2,\;\; A=\frac{1}{2H}\sinh Ht;\;\;\;\mbox{  for  }\Lambda=3H^2,\;\; A=\frac{1}{2H}\sin Ht.
\ee
%
We can deal simultaneously with both signs of $\Lambda$ by introducing
\[z=-2\dot{A}\left(\frac{3}{\Lambda}\right)^{1/2},\mbox{   so that   }\dot{z}=2A\left(\frac{\Lambda}{3}\right)^{1/2}=\left(1-\frac{\Lambda}{3}z^2\right)^{1/2}.\]
Set $C=F/A^2$ in (\ref{cc}) then
\[\left(1-\frac{\Lambda}{3}z^2\right)\frac{d^2F}{dz^2}+\frac{4\Lambda z}{3}\frac{dF}{dz}-\frac{4\Lambda}{3}F=0,\]
which is solved by
\[F=c_1z+c_2\left(z^4-\frac{18}{\Lambda}z^2-\frac{27}{\Lambda^2}\right).\]
Then with $B=G/A^2$, (\ref{bb}) becomes
\[\left(1-\frac{\Lambda}{3}z^2\right)\frac{d^2G}{dz^2}+\frac{4\Lambda z}{3}\frac{dG}{dz}-\frac{4\Lambda}{3}G=6C^2+\left(A\dot{C}-2\dot{A}C\right)^2.\]
with solution
\[G=PI+c_3z+c_4\left(z^4-\frac{18}{\Lambda}z^2-\frac{27}{\Lambda^2}\right),\]
where $PI$ is the particular integral. Again the metric is obtained subject to solving a second-order linear ODE.

For vacuum ($\Lambda=0$) we can obtain the general solution in (4-parameter) closed form:
\[A=\frac{t}{2},\;\;C=\frac{c_1}{t^2}+c_2t^4,\;\;B=\frac{4c_1^2}{t^6}-16c_1c_2+7c_2^2t^6+\frac{c_3}{t^2}+c_4t^4.\]
This is flat iff $B=C=0$ and it has Weyl tensor which is SD for $c_2=0=c_4$, or ASD for $c_1=0=c_3$.

\section{Other types}

 If the isometry group admits a 2-dimensional Abelian subgroup and the metric is orthogonally transitive (OT) then the methods of twistor theory can be used to find all Einstein solutions of any signature
  (see \cite{MW}, examples in \cite{CT1},\cite{CT2} and self--dual 
  Kahler examples in \cite{DP}). This will include all Bianchi types except for VIII and IX. Generically, there should be a reduction to a Painlev\'e
  equation (as happens in \cite{CT1} for the Bianchi III examples treated there, which are OT for the group generated by $<\partial_x,\partial_y>$)
  but type I, as seen above, is actually solvable in elementary functions (and the metrics found are all OT for the subgroup generated by $(\partial_y,\partial_z)$). It is possible to solve type II and obtain some
  examples which are not OT (and not in \cite{CT2}).
\subsection{Type II}
Consider the metric
\[g=dt^2+a^2\sigma_1^2+b^2\sigma_2^2+(f\sigma_2+c\sigma_3)^2,\]
with
\[d\sigma_1=\sigma_2\wedge\sigma_3,\;\;d\sigma_2=0=d\sigma_3,\]
which is therefore Bianchi type II. It is convenient to introduce coordinates $(x,y,z)$ by
\[\sigma_1=dx+ydz,\;\;\sigma_2=dy,\;\;\sigma_3=dz,\]
with corresponding Killing vectors
\[X_1=\partial_x,\;\;X_2=\partial_y-z\partial_x,\;\;X_3=\partial_z.\]
Abelian subgroups of the isometry group are generated by $<X_1,X_2>$ or $<X_1,X_3>$ but in both cases orthogonal transitivity implies diagonalisable, so that nondiagonalisable examples will not be OT.

For later use we note that the differential algebra of invariant 1-forms has the symmetry
\[ (\sigma_1,\sigma_2,\sigma_3)\rightarrow(\mu^{-1}\nu^{-1}\sigma_1,\mu^{-1}\sigma_2,\nu^{-1}\sigma_3)\]
for nonzero constants $\mu,\nu$ and under which the metric components change according to
\be\label{tr1}
(a,b,c,f)\rightarrow(\mu\nu a,\mu b,\nu c,\mu f).\ee
We exploit this symmetry below to fix some constants.

The Einstein equations are
\bea\label{eII}
\frac{\ddot{a}}{a}+\frac{\ddot{b}}{b}+\frac{\ddot{c}}{c}+2L^2&=&-\Lambda\nonumber\\
\frac{\ddot{a}}{a}+\frac{\dot{a}\dot{b}}{ab}+\frac{\dot{a}\dot{c}}{ac}-\frac{a^2}{2b^2c^2}&=&-\Lambda\nonumber\\
\frac{\ddot{b}}{b}+\frac{\dot{a}\dot{b}}{ab}+\frac{\dot{b}\dot{c}}{bc}+2L^2+\frac{a^2}{2b^2c^2}&=&-\Lambda\\
\frac{\ddot{c}}{c}+\frac{\dot{a}\dot{c}}{ac}+\frac{\dot{b}\dot{c}}{bc}-2L^2+\frac{a^2}{2b^2c^2}&=&-\Lambda\nonumber\\
\dot{L}+\left(\frac{\dot{a}}{a}+2\frac{\dot{c}}{c}\right)L&=&0,\nonumber
\eea
where
\be\label{l1}L=\frac12\left(\frac{\dot{f}}{b}-\frac{f\dot{c}}{cb}\right).\ee
The Hamiltonian constraint is
\be\label{h6}\frac{\dot{a}\dot{b}}{ab}+\frac{\dot{b}\dot{c}}{bc}+\frac{\dot{c}\dot{a}}{ca}-L^2+\frac{a^2}{4b^2c^2}+\Lambda=0.\ee
From the last of (\ref{eII}) deduce
\[L=\frac{c_1}{ac^2}.\]
In fact $c_1$ is one of the two twist potentials for the Abelian subgroup generated by $<X_1,X_3>$, the other being automatically zero. We insist that $c_1\neq 0$ since otherwise the metric is both OT and diagonalisable.

To proceed, change the time-coordinate by
\[dt=ac^2d\tau,\mbox{   so   }\frac{d}{d\tau}=ac^2\frac{d}{dt},\]
for then
\[\frac{d}{d\tau}\left(\frac{f}{c}\right)=2abcL=2c_1\left(\frac{b}{c}\right).\]
Next set $b=Xc$, eliminate $\Lambda$ between the third and fourth of (\ref{eII}) and substitute to find
\[\frac{\ddot{X}}{X}+\left(\frac{\dot{a}}{a}+2\frac{\dot{c}}{c} \right)\frac{\dot{X}}{X}+4L^2=0,\]
or equivalently
\[\frac{d^2X}{d\tau^2}+4c_1^2X=0,\]
which is readily solved. Under the symmetry (\ref{tr1}) we have $X\rightarrow \mu\nu^{-1}X$ so without loss of generality we may suppose that
\[X=\sin(2c_1\tau),\]
and the symmetry is reduced to (\ref{tr1}) but with $\mu=\nu$. Next the second of (\ref{eII}) translates to
\[\frac{d}{d\tau}\left(\frac{X}{a}\frac{da}{d\tau}\right)=\frac{a^4}{2X}-\Lambda a^2c^4X.\]
To make progress, set $\Lambda=0$, when
\[X^2\left(\frac{da}{d\tau}\right)^2=\frac14(a^6+c_2a^2),\]
for new constant $c_2$ which we suppose for now is nonzero and positive. The residual freedom in (\ref{tr1}) allows us to set $c_2=4$. Now
\[\frac{2da}{a(a^4+4)^{1/2}}=\frac{d\tau}{X}.\]
This can be simplified by setting $a^4=16g/(g-1)^2$ for then
\[g=c_3(\tan c_1\tau)^{-2/c_1}.\]
To find $c$ we go back to the Hamiltonian constraint and calculate
\[0=a^2c^4H=\left(\frac{c'}{c}+\frac{a'}{a}+\frac{X'}{2X}\right)^2-\frac14\left(\frac{X'}{X}\right)^2-c_1^2-\frac{1}{X^2},\]
with prime for $d/d\tau$. With $X$ known, this gives
\[acX^{1/2} =c_4(\tan c_1\tau)^{\beta/2c_1}\mbox{  where  }\beta^2=c_1^2+1.\]
This will give $c$ and therefore the general vacuum solution.
Solutions with $c_2<0$ can be obtained by analytic continuation and solutions with $c_2=0$ are SD. They have neutral signature, and are not cited in the list of type II solutions in \cite{CT2}.
The metric can be written in the form
\be\label{met55}
g=\frac{1}{8c_1^3}(v_0-v)e^{-2v}dv^2+\frac{2c_1}{v_0-v}\sigma_1^2+\frac{(v_0-v)}{2c_1}e^{-v}\left(\cosh v(\sigma_2^2+\sigma_3^2)+2\sinh v\sigma_2\sigma_3\right).\ee

\section*{Appendix}
Here we address the question of how many distinct ways there are to obtain real type VIII metrics by taking slices of diagonal complex type IX metrics. We shall see that the two
choices made in Section 3.1 are essentially all, up to equivalences.

With $\sigma_i,\Sigma_i$ the invariant 1-forms for type VIII and type IX respectively,
set $\sigma=(\sigma_1,\sigma_2,\sigma_3)^T$ and $\Sigma=(\Sigma_1,\Sigma_2,\Sigma_3)^T$. There is freedom
\[\sigma\rightarrow L\sigma,\;\;\Sigma\rightarrow P^T\Sigma\]
for real Lorentz transformation $L$ and complex orthogonal $P$. Suppose the forms are related by
\[\sigma=M\Sigma=\overline{M}\overline{\Sigma},\]
where the second equation is the reality condition. The allowed freedom has the effect
\[M\rightarrow LMP,\]
so, provided the top row of $M$ is not a null vector we can use complex $P$ to set the top row of $M$ to be $(\alpha,0,0)$ with $\alpha\neq 0$. (If the top row of $M$ is a null vector then we can use real $P$ to set it to be $a(1,i,0)$ for nonzero real $a$, in which case
\[\sigma_1=a(\Sigma_1+i\Sigma_2)\mbox{ when  }\sigma_1\wedge d\sigma_1=0,\]
which is a contradiction.)

Now
\[\sigma_1=\alpha\Sigma_1,\;\;\sigma_2=\beta\Sigma_1+\gamma\Sigma_2+\delta\Sigma_3,\;\;\sigma_3=\lambda\Sigma_1+\mu\Sigma_2+\nu\Sigma_3\]
for some $\beta,\gamma,\delta,\lambda,\mu,\nu$. From the exterior derivatives
\[d\sigma_1=\alpha d\Sigma_1=\alpha\Sigma_2\wedge\Sigma_3\]
but also
\[=2\sigma_2\wedge\sigma_3=2(\beta\Sigma_1+\gamma\Sigma_2+\delta\Sigma_3)\wedge(\lambda\Sigma_1+\mu\Sigma_2+\nu\Sigma_3).\]
Thus
\[\beta\nu-\lambda\delta=0=\beta\mu-\lambda\gamma,\;\;2(\gamma\nu-\delta\mu)=\alpha,\]
 whence $\beta=0=\lambda$.
With $M$ parametrised as
\[M=\left(\begin{array}{cc}
           \alpha & 0\\
          0 & \tilde{M}\\
\end{array}\right),\]
parametrise $\tilde{M}$ as
\[\left(\begin{array}{cc}
        \beta & \gamma\\
           \delta & \epsilon\\
 \end{array}\right)\]
 then
 \[d\sigma_2=\beta d\Sigma_2+\gamma d\Sigma_3=\beta\Sigma_3\wedge\Sigma_1+\gamma\Sigma_1\wedge\Sigma_2\]
  but also
  \[= -\sigma_3\wedge\sigma_1=-\alpha(\delta\Sigma_2+\epsilon\Sigma_3)\wedge\Sigma_1\]
 so that
 \[\alpha\delta=\gamma,\;\;\alpha\epsilon=-\beta.\]
 Consideration of $d\sigma_3$ similarly leads to
 \[\alpha\beta=\epsilon,\;\;\alpha\gamma=-\delta,\]
 so that, since e.g. $\beta$ and $\gamma$ are not both zero, we deduce that $\alpha^2=-1$. Without loss of generality we may choose $\alpha=i$ whence also
 \[\epsilon=i\beta,\;\;\delta=-i\gamma.\]
 The remaining condition from consideration of $d\sigma_1$ now entails
 \be\label{A1}\beta^2+\gamma^2=1.\ee
%
%
%
The complex type IX metric is
\[g=dt^2+A\Sigma_1^2+B\Sigma_2^2+C\Sigma_3^2,\]
and we wish to obtain a real type VIII metric as
\[=dt^2-A\sigma_1^2+B(\beta\sigma_2+i\gamma\sigma_3)^2+C(\gamma\sigma_2-i\beta\sigma_3)^2\]
\[=dt^2-A\sigma_1^2+(B\beta^2+C\gamma^2)\sigma_2^2+2i\beta\gamma(B-C)\sigma_2\sigma_3-(B\gamma^2+C\beta^2)\sigma_3^2.\]
For this to be real we require the following to be real
\[A,\;\;B\beta^2+C\gamma^2,\;\;B\gamma^2+C\beta^2,\;\;i\beta\gamma(B-C).\]
By taking combinations of the last three (first minus second plus or minus twice the third) we see that $(\beta\pm i\gamma)^2(B-C)$ must be real and therefore so must $(\beta+ i\gamma)^2/(\beta- i\gamma)^2$, so that
\[(\beta+ i\gamma)^2(\overline{\beta}+ i\overline{\gamma})^2=(\beta- i\gamma)^2(\overline{\beta}- i\overline{\gamma})^2\]
 whence
 \be\label{k1} (\beta\overline{\beta}-\gamma\overline{\gamma})(\gamma\overline{\beta}+\beta\overline{\gamma})=0.\ee
 Note that
 \[(\beta\overline{\beta}-\gamma\overline{\gamma})^2+(\gamma\overline{\beta}+\beta\overline{\gamma})^2=1\]
 by virtue of (\ref{A1}) so that there is no loss of generality in supposing that whichever factor is zero in (\ref{k1}) the other factor can be assumed to be one.

 The reality conditions on the $\Sigma_i$ are
 \[\overline{\Sigma}_1=-\Sigma_1,\;\;\beta\Sigma_2+\gamma\Sigma_3=\overline{\beta}\overline{\Sigma}_2+\overline{\gamma}\overline{\Sigma}_3,\;\;
  -i\gamma\Sigma_2+i\beta\Sigma_3=i\overline{\gamma}\overline{\Sigma}_2-i\overline{\beta}\overline{\Sigma}_3,\]
whence also
\[\Sigma_2=(\beta\overline{\beta}-\gamma\overline{\gamma})\overline{\Sigma}_2+(\gamma\overline{\beta}+\beta\overline{\gamma})\overline{\Sigma}_3,\;\;
 \Sigma_3=(\gamma\overline{\beta}+\beta\overline{\gamma})\overline{\Sigma}_2-(\beta\overline{\beta}-\gamma\overline{\gamma})\overline{\Sigma}_3.\]
Now by(\ref{k1}) these simplify and we have just two choices: either
\[\Sigma_2=\overline{\Sigma}_2,\;\;\Sigma_3=-\overline{\Sigma}_3\]
which is Kamada's choice (\ref{s1}) or
\[\Sigma_2=\overline{\Sigma}_3\]
which is (\ref{ch1}), the other choice.

\end{document}